\documentclass{article}
\usepackage{spconf,amsmath,graphicx}
\usepackage{subfigure}
\usepackage{multirow}
\usepackage{booktabs}
\usepackage{amssymb}
\usepackage{lipsum}
 
\newcommand\blfootnote[1]{%
  \begingroup
  \renewcommand\thefootnote{}\footnote{#1}%
  \addtocounter{footnote}{-1}%
  \endgroup
}

\usepackage{hyperref}

\hypersetup{hidelinks,
	colorlinks=true,
	allcolors=black,
	pdfstartview=Fit,
	breaklinks=true}

\title{Mingling or Misalignment? Temporal Shift for Speech Emotion Recognition with Pre-trained Representations}
\name{Siyuan Shen$^1$ \qquad Feng Liu$^1$ \qquad Aimin Zhou$^{1,2,\star}$ }

\address{$^1$East China Normal University, Shanghai, China \\
	     $^2$Shanghai Institute of AI for Education, Shanghai, China}
\begin{document}
%\ninept
%
\maketitle
\begin{abstract}
    Fueled by recent advances of self-supervised models, pre-trained speech representations proved effective for the downstream speech emotion recognition (SER) task. Most prior works mainly focus on exploiting pre-trained representations and just adopt a linear head on top of the pre-trained model, neglecting the design of the downstream network. In this paper, we propose a temporal shift module to mingle channel-wise information without introducing any parameter or FLOP. With the temporal shift module, three designed baseline building blocks evolve into corresponding shift variants, i.e. ShiftCNN, ShiftLSTM, and Shiftformer. Moreover, to balance the trade-off between mingling and misalignment, we propose two technical strategies, placement of shift and proportion of shift. The family of temporal shift models all outperforms the state-of-the-art methods on the benchmark IEMOCAP dataset under both finetuning and feature extraction settings. Our code is available at \href{https://github.com/ECNU-Cross-Innovation-Lab/ShiftSER}{https://github.com/ECNU-Cross-Innovation-Lab/ShiftSER}.
\end{abstract}
\section{Introduction}

Speech emotion recognition (SER) refers to recognizing human emotion and affective states from audio. Benefiting from different emotional datasets annotated with categorical or dimensional labels, deep neural networks (DNNs) have shown great capability of recognizing emotion from speech efficiently \cite{wu2022neural}. The standard practice is training from scratch with conventional features \cite{rajamani2021novel}. However, the relatively small SER datasets \cite{busso2008iemocap}, containing only a few speakers and utterances, and conventional hand-crafted acoustic features restrict the performance \cite{chen2022wavlm}.\blfootnote{\ This paper is funded by  Special Project on Support for Regulative Intelligence Technology under the Shanghai Science and Technology Innovation Plan 2022 on "Research on the Theory, Assessment System and Prototype System for Enhancing Emotional Cognition in the General Population" (Grant No. 22511105901).\\
$^{\star}$ Corresponding author. amzhou@cs.ecnu.edu.cn}

Inspired by the success of pre-trained features like wav2vec 2.0 \cite{baevski2020wav2vec} and HuBERT \cite{hsu2021hubert} in other speech tasks, some researchers begin to validate its superiority over hand-engineered features in SER \cite{xia2021temporal}. Some works present fusion methods of pre-trained and traditional features \cite{pepino2021emotion} while others explore task adaptive pre-training strategies for SER \cite{chen2021exploring}. However, most of previous works mainly focus on exploiting pre-trained representations and just adopt linear head on top of the pre-trained model, neglecting the design of downstream network. The building of specialist models for speech downstream tasks is also necessary for pre-trained representations.

\begin{figure}[!t]
    \centering
    \subfigure[Origin.]{
          \includegraphics[width=0.33\linewidth]{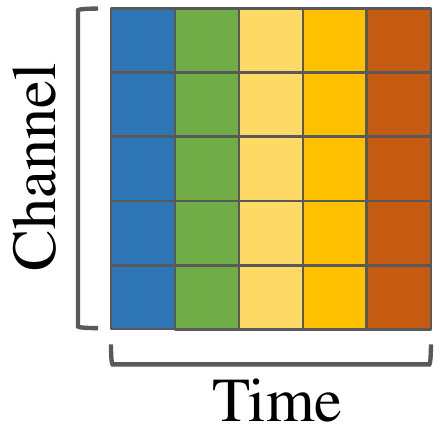}
            %\caption{Original.}
            \label{fig1:a}
    }
    \hfil
    \subfigure[Unidirection.]{
          \includegraphics[width=0.25\linewidth]{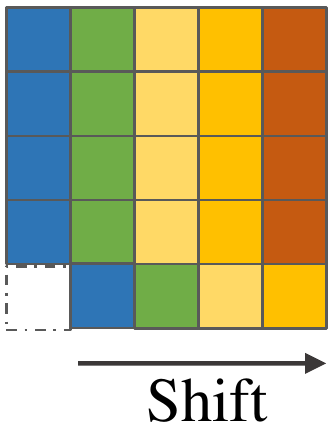}
            %\caption{Unidirection.}
            \label{fig1:b}
    }
    \hfil
    \subfigure[Bidirection.]{
          \includegraphics[width=0.25\linewidth]{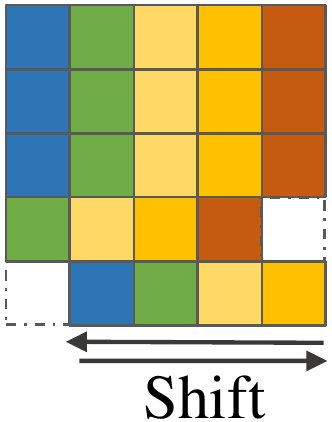}
            %\caption{Bidirection.}
            \label{fig1:c}
    }
    \caption{The original speech representation and temporal shift.}
    \label{fig1}
\end{figure}

Therefore, we take initiatives to explore the architecture of the networks for SER on pre-trained representations and propose a parameter-free, FLOP-free temporal shift module to promote channel mingling. We first draw inspirations from the advancement of Transformers \cite{vaswani2017attention} as well as modern CNNs \cite{liu2022convnet} and investigate the common network configurations to set strong baselines. Then, as shown in Figure \ref{fig1}, we introduce an efficient channel-wise network module, where partial channels are shifted along the temporal dimension by one frame to mingle the past feature to the present one. The core idea of our temporal shift is to introduce channel-wise information mingling. It is noteworthy that the features are shifted partially and thus endows the whole model with partial receptive field enlargement, different from the sufficient frame-level receptive field enlargement by stacking pretrained CNN or Transformer blocks. Moreover, our temporal shift serves as the network module similar to the shift operations in computer vision \cite{lin2019tsm} rather than the common augmentation applications in audio processing \cite{wang22interspeech}. However, such channel-wise partial shift seems to go against the characteristic of alignment in many speech tasks \cite{graves2012sequence}.  From the perspective of alignment, the information contained in the shifted channels becomes inaccessible for the original frame, indicating misalignment along with mingling. To balance the trade-off between mingling and misalignment, we propose two strategies for applying temporal shift, including proportion and placement. Proportion of shift is defined to adjust the ratio of shifted channels and the placement of shift can be unified into two manners (Figure \ref{fig3}). To the best of our knowledge, we are the first to propose channel-wise shift operation in audio processing and apply temporal shift to three mainstream building blocks. The family of shift models, including ShiftCNN, Shiftformer and ShiftLSTM, all outperform the state-of-the-art methods on the benchmark IEMOCAP dataset under both finetuning and feature extraction settings.

\section{Methodology}
In this section, we first illustrate the property of temporal shift and its application strategy. Then we specify the design of basic building blocks and the shift variants.

\subsection{Temporal Shift}
As motivated, our temporal shift is to introduce channel-wise information mingling to the standard building block. Part of the channels will be shifted to its neighbouring time stamp and the rest channels remain as shown in Figure \ref{fig1}. The unidirectional shift mingles the past feature to the current one while the bidirectional shift combines the future feature with the current one additionally.

Nonetheless, the partial shifted data leads to frame misalignment and the original information contained in the shifted channels is inaccessible for that time stamp. Specifically, for each time stamp, the un-shifted features are forced to adapt to the neighboring shifted channels while the shifted channels are moved aside from the position it used to be. To balance the trade-off between mingling and misalignment, we provide two strategies for temporal shift.

\textbf{Proportion of shift.} The proportion of shifted channels can be defined as the hyperparameter $\alpha$, adjusting the bargain between mingling and misalignment. Intuitively, if $\alpha$ is kept small enough, the negative effect can be negligible. If $\alpha$ is set large, the information mingling is promoted but the underlying misalignment may cause performance degradation. We will measure the variants with shift proportion in $\left \{ 1/2,1/4,1/8,1/16 \right \}$.

\begin{figure}[!t]
    \centering
    \subfigure[In-place Shift.]{
          \includegraphics[width=0.55\linewidth]{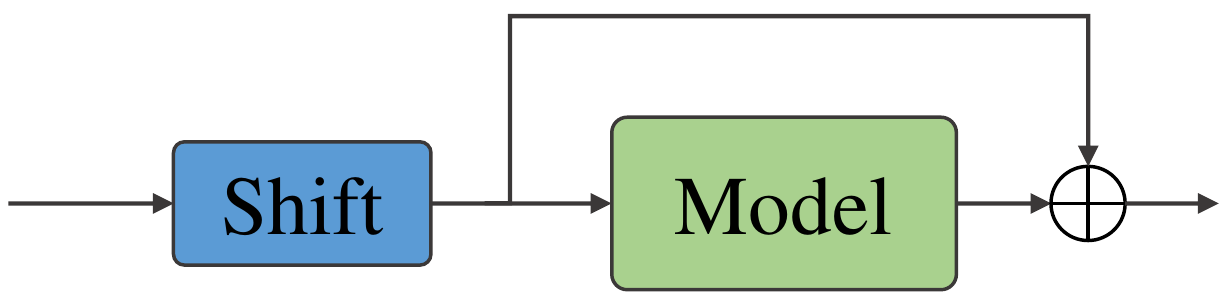}
          \label{fig3:a}
    }
    \subfigure[Residual Shift.]{
          \includegraphics[width=0.38\linewidth]{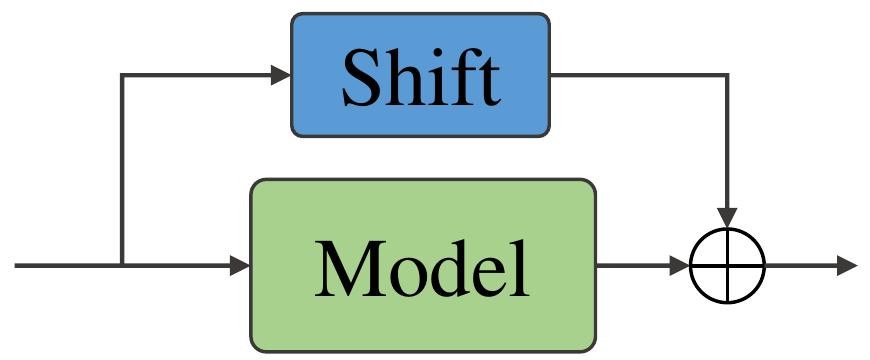}
            \label{fig3:b}
    }
    \caption{Two types of the building blocks of our temporal-shift networks, namely in-place shift and residual shift.}
    \label{fig3}
\end{figure}

\textbf{Placement of shift.} Exploring the proper location of temporal shift in the neural network is essential. As shown in Figure \ref{fig3}, The manner of applying temporal shift to the model can be unified in two ways, including residual shift and in-place shift. In-place shift (Figure \ref{fig3:a}) forces the model to fit the shifted data, where the effects of both sides are great. Residual shift (Figure \ref{fig3:b}) is placed on the branch of the network, maintaining both the aligned data and misaligned data. The shifted representations from the residual connection can serve as a complementary to the original representations, behaving like an \emph{ensemble} \cite{veit2016residual}.

In addition to technical strategies, the underlying misalignment can be remitted by the SER task from pre-trained representations itself. Specifically, the characteristic of SER task indicates its focus on invariant and discriminative representations rather than alignments between the input and target sequence in other speech tasks \cite{graves2012sequence}.

\subsection{Specification of Temporal Shift Models}
We insert temporal shift module into three types of popular building blocks in SER, including convolution neural network, recurrent neural network, and Transformer. The number of channels $C$ matches the representations of pre-trained wav2vec 2.0 Base and HuBERT Base. The number of blocks $B$ is set to ensure the number of parameters roughly the same (9.5M). It is worth noting that our temporal shift can be inserted flexibly and the proportion of shift can be adjusted as the hyperparameter. For uniformity, we adopt the same architecture with fixed proportion in our experiments to validate its effectiveness. The isotropic architectures are summarized as follows.

\begin{itemize}
\item ShiftCNN: $C=(768,3072,768), B=2, \alpha=1/16$
\item Shiftformer: $C=(768,3072,768), B=2, \alpha=1/4$
\item ShiftLSTM: $C=(768,1536), B=1, \alpha=1/4$
\end{itemize}

\begin{figure}[t]
    \centering
    \subfigure[CNN]{
          \includegraphics[width=0.25\linewidth]{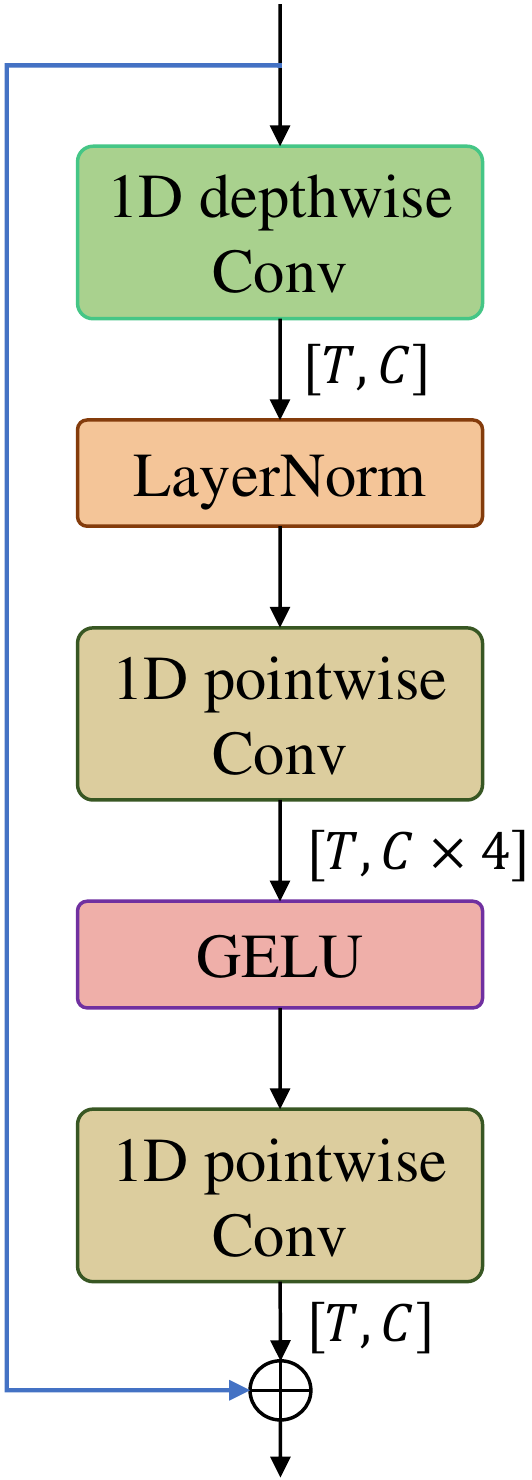}
          \label{figtrans:a}
    }
    \hspace{0.03\linewidth}
    \subfigure[Transformer]{
          \includegraphics[width=0.245\linewidth]{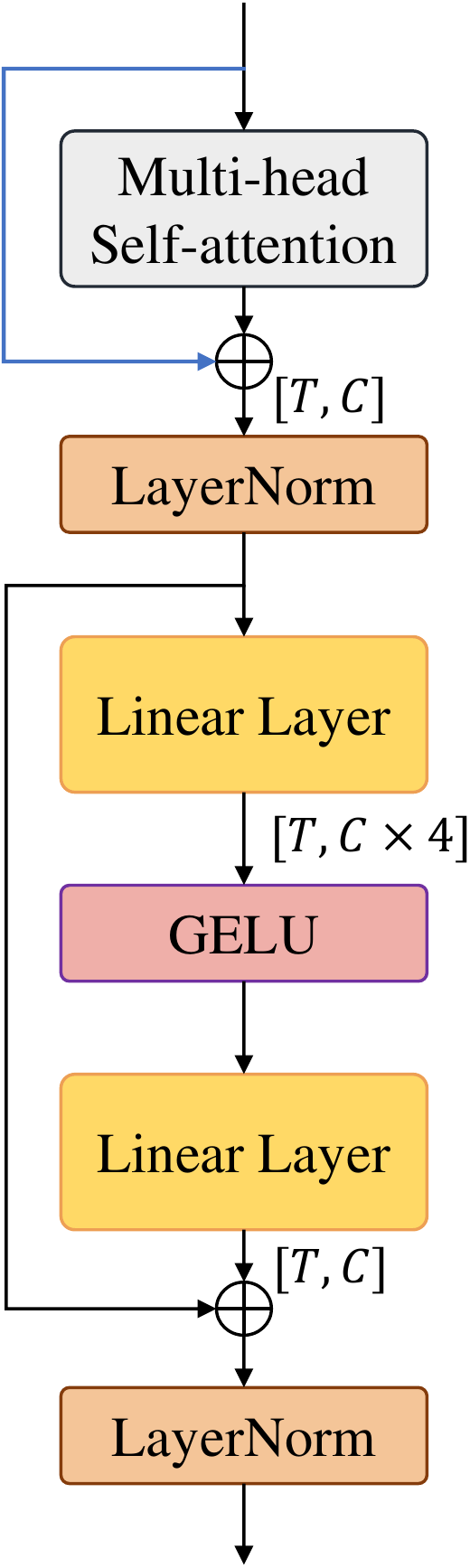}
          \label{figtrans:b}
    }
    \hspace{0.03\linewidth}
    \subfigure[Shiftformer]{
          \includegraphics[width=0.235\linewidth]{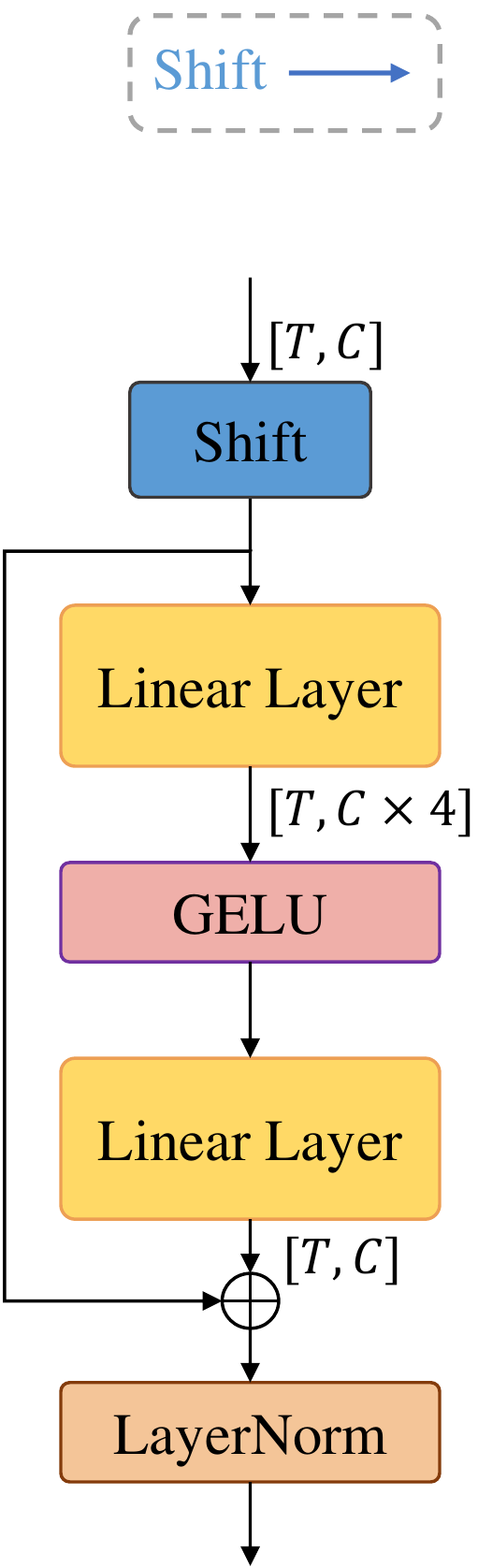}
            \label{figtrans:c}
    }
    \caption{Block design of CNN, Transformer, and Shiftformer. $T$ and $C$ denote temporal and channel dimensions of feature maps respectively. Temporal shift can be inserted flexibly and our choice of residual shift is colored blue.}
    \label{figtrans}
\end{figure}

\textbf{CNN.} We draw inspirations from the advancement of ConvNext blocks \cite{liu2022convnet} and time-channel separable convolution. This pure convolution neural network architecture described in Figure \ref{figtrans:a} can serve as a strong baseline. For its shift variant ShiftCNN with kernel size 7, we have the last block with residual shift for 1/16 proportion, which enables channel-aligned modeling capability and subtle channel-wise information fusion.

\textbf{Transformer.} We employ the standard Transformer with relative positional encoding (RPE) as shown in Figure \ref{figtrans:b}. However, the standard Transformer with residual shift brings no more improvement (Table \ref{tableab} residual$\rightarrow$shift of Transformer). Inspired by the recent literature \cite{yu2022metaformer}, we further sidestep self-attention and view Transformer as a more general architecture, consisting of token mixer and MLP. The building of Shiftformer (Figure \ref{figtrans:c}), closer to MLP-like architecture, is based on the replacement of self-attention with temporal shift as token mixer, involving a more radical bidirectional shift manner. Our shiftformer enjoys effectiveness of the general Transformer architecture as well as the channel-wise mingling of temporal shift.

\textbf{RNN.} Bidirectional long short-term memory (LSTM) \cite{hochreiter1997long} is set to capture both past and future contexts. ShiftLSTM is constructed by straightforward in-place shift with moderate 1/4 shift proportion.

\section{Experiments}
In this section, we first observe the main properties of temporal shift module. Next we demonstrate the superiority of the family of shift models on the benchmark dataset IEMOCAP under both finetuning and feature extraction settings. Finally, we ablate key components of our proposed models.

\subsection{Experimental Setup}
\textbf{Dataset.} Interactive emotional dyadic motion capture database (IEMOCAP) \cite{busso2008iemocap} is a widely-used benchmark SER dataset with total length of about 12h, containing five sessions with one male and one female each. We adopt leave-one-session-out 5-fold cross-validation and four emotion categories, including 1636 happy utterances, 1084 sad utterances, 1103 angry utterances, and 1708 neutral utterances. Each sample is clipped to 7.5 seconds and the sampling rate is 16kHz.

\textbf{Evaluation.} We do comparisons under two configurations, feature extraction and finetuning. Feature extraction \cite{pepino2021emotion} is to evaluate the sole downstream model, where the frozen pre-trained model is taken as the feature extractor and the downstream models are trained in the supervised manner. And under the finetuning configuration, the parameters of both upstream pre-trained models (only the Transformer layers) and downstream models are updated during training. 

The unweighted accuracy (UA) and weighted accuracy (WA) are used as evaluation metrics. UA is computed as the average over individual class accuracies while WA is the correct prediction over the total samples. The reported accuracy is the average of 5 folds results.

\textbf{Implementation Details.} We conduct experiments with two representative self-supervised pre-trained models, wav2vec 2.0 Base \cite{baevski2020wav2vec} and HuBERT Base \cite{hsu2021hubert}. We finetune the models using Adam optimizer with learning rate of $5\times10^{-4}$. We use AdamW optimizer with cosine scheduler and decay rate of 0.1 for feature extraction. We train at batch size of 32 for 100 epochs with 5 epochs of linear warm-up.

\begin{figure}[!t]
    \centering
    \subfigure[]{
          \includegraphics[width=0.47\linewidth]{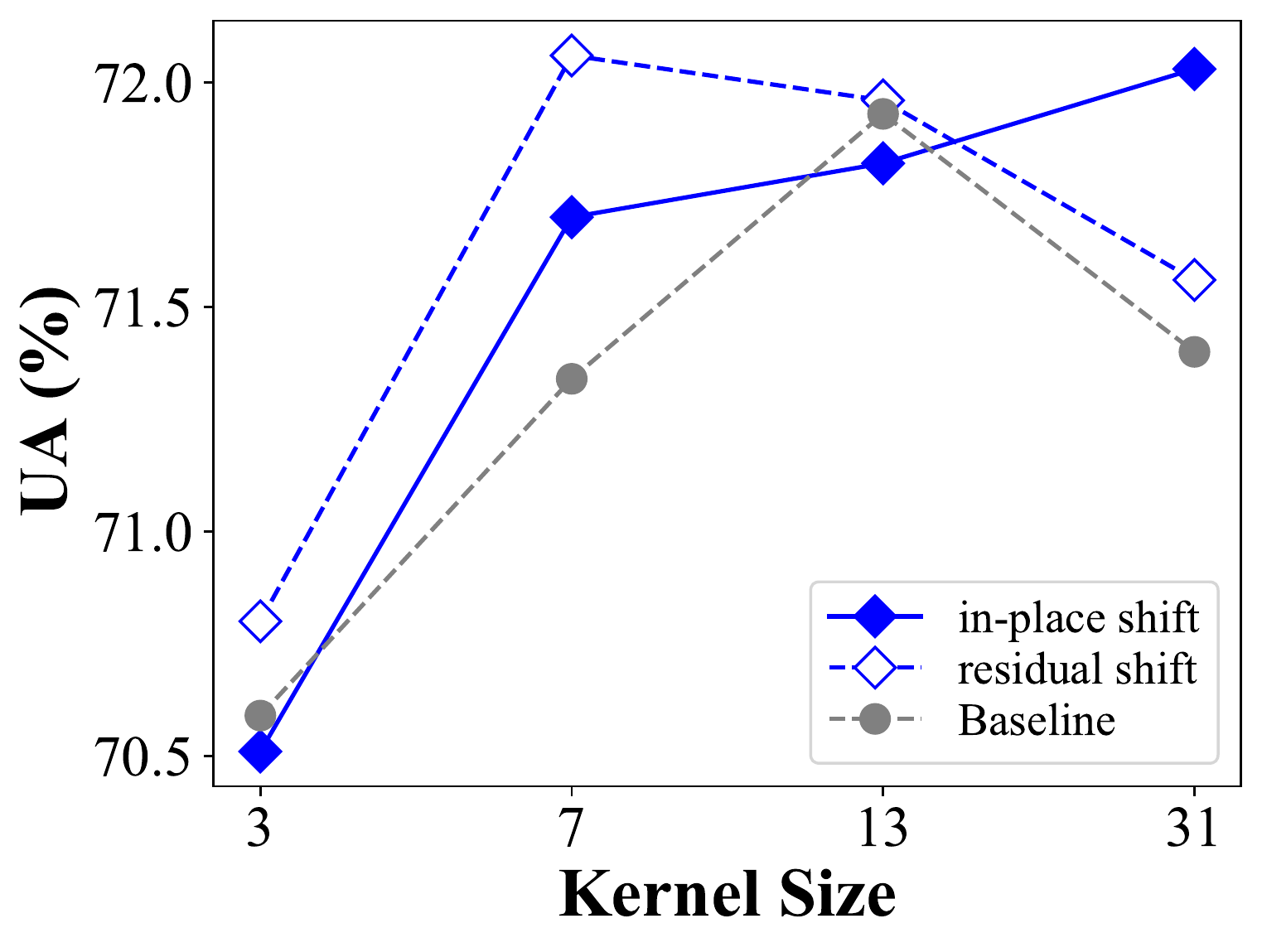}
          \label{figcrnn:a}
    }
    \hfil
    \subfigure[]{
          \includegraphics[width=0.47\linewidth]{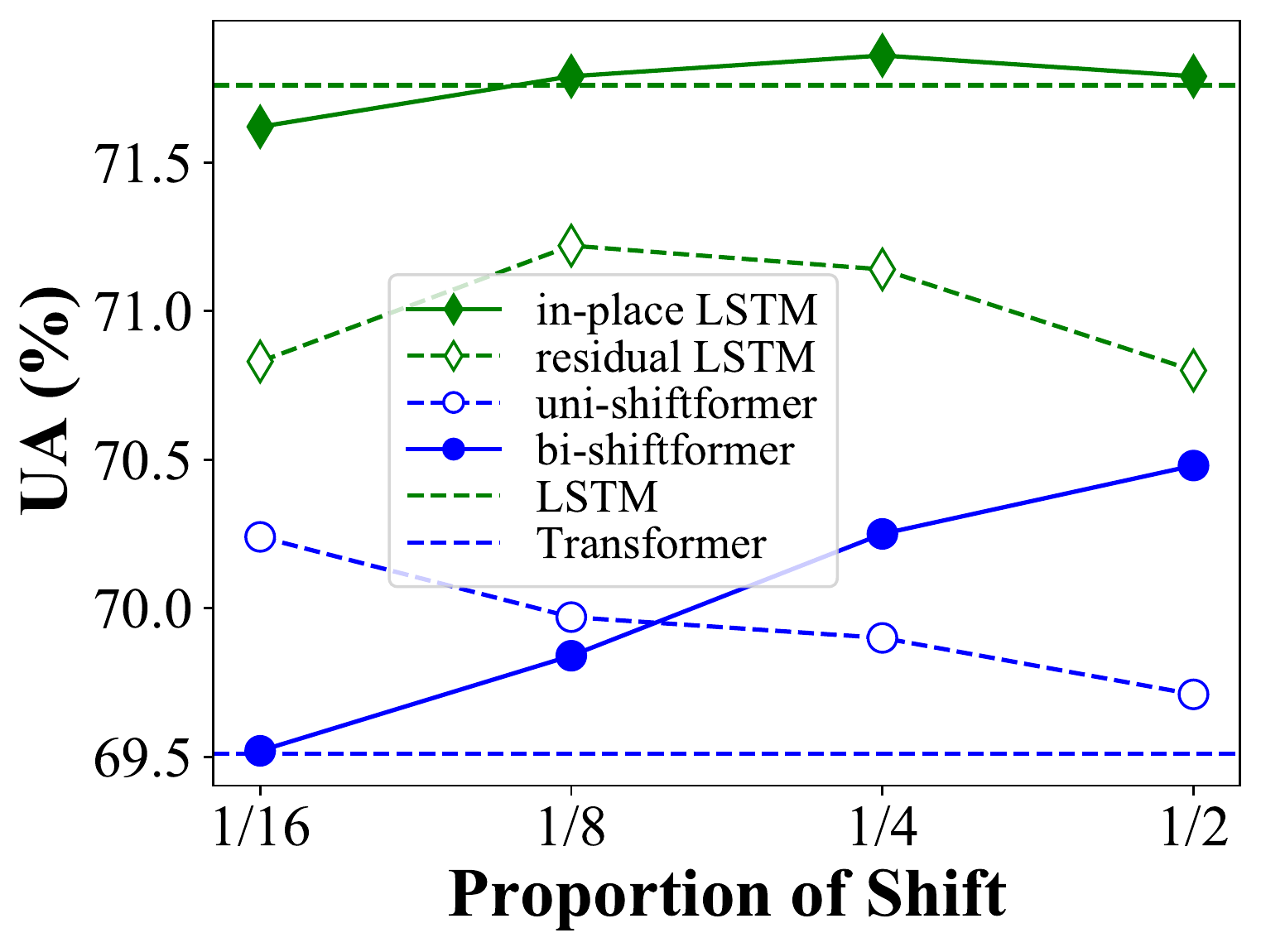}
            \label{figcrnn:b}
    }
    \caption{(a) The impact of temporal shift on ShiftCNN as kernel size varies. (b) The impact of shift proportion on ShiftLSTM and Shiftformer.}
    \label{figcrnn}
\end{figure}

\subsection{Properties of shift variants}
We conduct experiments on HuBERT feature extraction and observe the behavior of temporal shift. Since our temporal shift module provides channel-wise enlargement of receptive field, the effect is closely related to that of the baseline model. Figure \ref{figcrnn:a} varies CNN kernel sizes to probe the potential of temporal shift, where residual shift brings more gain with small kernels (from 71.3\% to 72.1\% for size 7) than large ones. Residual shift also achieves better performance than in-place shift for all relatively small kernels. From similar perspective as \cite{veit2016residual}, the residual shift makes the model an implicit \emph{ensemble} composed of shift and origin modules, thus benefiting from both channel-aligned and channel-mingled information at no extra computation. However, the projection shortcut added to bidirectional LSTM in Figure \ref{figcrnn:b} introduces extra parameters and puts residual shift at a disadvantage, leading to nearly 1\% degradation. The accuracy of Shiftformer increases steadily with proportion of bidirectional shift in Figure \ref{figcrnn:b}, surpassing the standard Transformer by about 1\% with 27\% fewer parameters. This suggests high proportion with bidirectional manner improves temporal modeling for MLP-like architecture and meanwhile avoids high complexity of self-attention operation.

\subsection{Comparison with State-of-the-Arts}
\textbf{Baselines.} We compare our proposed models with existing state-of-the-art methods under both finetuning and feature extraction settings. Finetuning methods include vanilla finetuning, P-TAPT, TAPT \cite{chen2021exploring}, CNN-BiLSTM \cite{xia2021temporal} and SNP, TAP \cite{gat2022speaker}. Feature extraction methods include conventional feature (AR-GRU \cite{rajamani2021novel}, SeqCap with NAS \cite{wu2022neural}), feature fusion (eGeMAPS+wav2vec 2.0 \cite{pepino2021emotion}, MFCC+Spectrogram+W2E \cite{zou2022speech}) and pre-trained features (wav2vec 2.0 \cite{baevski2020wav2vec}, HuBERT \cite{hsu2021hubert} and WavLM \cite{chen2022wavlm}). Moreover, we adopt temporal shift as data augmentation on hidden states for comparison, namely employed randomly just in the training stage.

\begin{table}[t]
\centering
\small
\begin{tabular}{@{}l|ll@{}}
\toprule
\textbf{Type}                                                                       & \textbf{Method}     & \textbf{UA (\%)} \\ \midrule
\multirow{4}{*}{Baseline}                                                           & CNN-BiLSTM          & 66.9             \\
                                                                                    & Vanilla Finetuning  & 69.9             \\
                                                                                    & P-TAPT              & 74.3             \\
                                                                                    & TAP+HuBERT Large    & 74.2             \\ \midrule
\multirow{3}{*}{\begin{tabular}[c]{@{}l@{}}Temporal\\ Shift\\ Augment\end{tabular}} & LSTM+Augment        & 74.1             \\
                                                                                    & Transformer+Augment & \textbf{74.8}    \\
                                                                                    & CNN+Augment         & 74.3             \\ \midrule
\multirow{3}{*}{\begin{tabular}[c]{@{}l@{}}Temporal\\ Shift\end{tabular}}           & ShiftLSTM (ours)    & 74.5             \\
                                                                                    & Shiftformer (ours)  & 74.7             \\
                                                                                    & ShiftCNN (ours)     & \textbf{74.8}    \\ \bottomrule
\end{tabular}
\caption{Comparison of UA with finetuning methods on IEMOCAP. Results are from original paper.}
\label{tableFinetune}
\end{table}

\begin{table}[t]
\centering
\small
\begin{tabular}{@{}l|l|cc@{}}
\toprule
\textbf{Feature} & \textbf{Method}     & \textbf{WA (\%)} & \textbf{UA (\%)} \\ \midrule
\multirow{2}{*}{Convention}           & AR-GRU              & 66.9               & 68.3               \\
                                      & SeqCap with NAS     & 70.5               & 56.9               \\ \midrule
\multirow{2}{*}{Fusion}               & eGeMAPS+wav2vec 2.0 & -                  & 66.3               \\
                                      & MFCC+Spec.+w2E      & 69.8               & 71.1               \\ \midrule
\multirow{6}{*}{Pre-trained}          & wav2vec 2.0 Large   & 65.6               & -                  \\
                                      & HuBERT Large        & 67.6               & -                  \\
                                      & WavLM Large         & 70.6               & -                  \\
                                      & ShiftCNN (ours)            & 71.9               & \textbf{72.8}      \\
                                      & ShiftLSTM (ours)           & 69.8               & 70.6               \\
                                      & Shiftformer (ours)         & \textbf{72.1}      & 72.7               \\ \bottomrule
\end{tabular}
\caption{Comparison with feature extraction on IEMOCAP. The results of pre-trained models are cited from \cite{chen2022wavlm} and others are from original paper. To be consistent with prior works, we adopt wav2vec 2.0 Base features. }
\label{tableFeatureEX}
\end{table}

\begin{table}[t]
\centering
\small
\begin{tabular}{@{}l|l|cc@{}}
\toprule
\textbf{Ablation}            & \textbf{Variant}        & \textbf{FeatEx.} & \textbf{Finetune} \\ \midrule
\multirow{4}{*}{CNN}         & baseline                & 71.3          & 74.1          \\
                             & conv$\rightarrow$depthwise          & 68.7          & 74.2              \\
                             & LN$\rightarrow$BN                   & 70.2          & 73.7              \\
                             & \textbf{residual$\rightarrow$shift} & \textbf{71.7} & \textbf{74.8} \\ \midrule
\multirow{4}{*}{Transformer} & baseline                & 69.5          & 74.5          \\
                             & RPE$\rightarrow$APE                 & 61.6          & 73.7          \\
                             & residual$\rightarrow$shift            & 69.3          & 74.5          \\
                             & MHSA$\rightarrow$pooling            & 69.5          & 73.7          \\
                             & \textbf{MHSA$\rightarrow$shift}     & \textbf{70.3} & \textbf{74.7} \\ \midrule
\multirow{2}{*}{LSTM}        & baseline                & 71.8          & 74.3          \\
                             & \textbf{identity$\rightarrow$shift} & \textbf{71.9} & \textbf{74.5} \\ \bottomrule
\end{tabular}
\caption{Ablation study for components of building blocks.}
\label{tableab}
\end{table}

\textbf{Comparison on finetuning.} As shown in Table \ref{tableFinetune}, the family of temporal shift models outperforms the state-of-the-art finetuning methods. For the adaptive finetuning methods, TAPT requires another Hubert-like pre-training stage on IEMOCAP while TAP utilizes a more advanced pre-trained model HuBERT Large, but still falling behind ours trained solely on wav2vec 2.0 Base by 0.6 \%. Notably, our temporal shift taken as data augmentation also shows competitive performance, signifying the channel-wise mingling mechanism and our strong baseline building blocks.

\textbf{Comparison on feature extraction.} Table \ref{tableFeatureEX} is split into multiple parts to include methods adopting different types of features. We follow the evaluation protocol as \cite{pepino2021emotion}, namely adopting the trainable weighted sum of embeddings. Taking the wav2vec 2.0 Base as the feature extractor, our ShiftCNN and Shiftformer outperform all the other methods, even attaining better performance than one of the latest advanced pre-trained model WavLM Large by 1.5\%.

\subsection{Ablation Studies}
We conduct ablation studies on CNN, Transformer, and LSTM respectively in Table \ref{tableab} under both wav2vec 2.0 finetuning and HuBERT feature extraction (dubbed FeatEx.). UA is reported in the table. The advancement of the overall architecture is verified by the ablation of key components, covering types of normalization, convolution and position encoding. Interestingly, Transformer with residual shift fails to catch up with the standard one while our shiftformer, replacing multi-head self-attention (MHSA) with shift operation, outperforming the others with 3.6M fewer parameters. This indicates the benefit of reducing complexity and introducing channel-wise information mingling.

\section{Conclusion}
In this paper, we propose temporal shift module to introduce channel-wise mingling for downstream models. We hope our shift variants inspire future research on downstream model design and look forward further application of temporal shift.

\vfill\pagebreak

\bibliographystyle{IEEEbib}
\bibliography{main}

\end{document}